\documentclass[prc,showpacs,showkeys,superscriptaddress,nofootinbib,twocolumn,floatfix]{revtex4}
\usepackage{amsfonts}
\usepackage{graphicx,amsmath,amssymb}

\def\blfootnote{\xdef\@thefnmark{}\@footnotetext}

\newcommand{\lsim}{\lesssim}

\newcommand{\ket}{K\! E_T}

\begin{document}

\title{Scaling of Elliptic Flow, Recombination and Sequential Freeze-Out
of Hadrons in Heavy-Ion Collisions}

\author{Min He}
\affiliation{Cyclotron Institute and Department of Physics and Astronomy, Texas A\&M University, College Station, TX 77843, USA}

\author{Rainer J.\ Fries}
\affiliation{Cyclotron Institute and Department of Physics and Astronomy, Texas A\&M University, College Station, TX 77843, USA}
\affiliation{RIKEN/BNL Research Center, Brookhaven National Laboratory,
Upton, NY 11973, USA}

\author{Ralf Rapp}
\affiliation{Cyclotron Institute and Department of Physics and Astronomy, Texas A\&M University, College Station, TX 77843, USA}

\date{\today}

\begin{abstract}
  The scaling properties of elliptic flow of hadrons produced in
  ultrarelativistic heavy-ion collisions are investigated at low transverse
  momenta, $p_T\lsim 2$\,GeV. Utilizing
  empirical parameterizations of a thermalized fireball with
  collective-flow fields, Resonance Recombination Model (RRM) is employed
  to describe hadronization via quark coalescence at the hadronization
  transition. We reconfirm that RRM converts equilibrium quark distribution
  functions into equilibrated hadron spectra including the effects of
  space-momentum correlations on elliptic flow. This provides the basis
  for a controlled extraction of quark distributions of the bulk matter
  at hadronization from spectra of multi-strange hadrons which are believed
  to decouple close to the critical temperature. The resulting elliptic
  flow from empirical fits at RHIC exhibits transverse kinetic-energy and
  valence-quark scaling. Utilizing the well-established concept of
  sequential freeze-out we find that the scaling at low momenta can be
  extended to bulk hadrons ($\pi$, $K$, $p$) at thermal freeze-out, and
  thus result in an overall description compatible with both equilibrium
  hydrodynamics and quark recombination.

\end{abstract}

\maketitle

\section{Introduction}
\label{sec_intro}
Ultrarelativistic heavy-ion collisions (URHICs) enable the creation and
study of superdense strongly interacting matter, possibly associated
with the formation of novel phases where partons deconfine and chiral
symmetry is restored~\cite{RHIC2005}.
One major finding by the experimental program at the Relativistic
Heavy Ion Collider (RHIC) at Brookhaven National Laboratory is the
large azimuthal anisotropy of hadron transverse-momentum ($p_T$)
spectra in non-central nuclear collisions, the so-called elliptic
flow~\cite{v2review2009}. It is quantified by the second harmonic
coefficient, $v_2(p_T)$, in the Fourier expansion of the azimuthal
angle, $\phi$, for the $p_T$-spectra of hadrons.

Experimental measurements suggest that the behavior of $v_2(p_T)$
may be classified into three regimes. At high transverse momentum,
$p_T\gtrsim 6$\,GeV, the azimuthal anisotropy is believed to be due
to the path-length dependence of high-energy partons as they
traverse the hot medium and lose energy. We will not be concerned
any further with this mechanism in the present work.  At low
transverse momentum, $p_T\lesssim 2$\,GeV, $v_2$ increases with
$p_T$ essentially linearly but with a delayed onset for hadrons with
increasing mass, giving rise to the so-called mass ordering of
elliptic flow~\cite{RHICv2}. This regime is well described by
hydrodynamic simulations and thus indicates a large degree of
equilibration, encompassing more than 90\% of the observed
particles~\cite{shuryak,Heinz20032004,Heinz2000,Heinz2001,Hirano2002}.
At intermediate transverse momenta, $2$\,GeV$\lesssim p_T\lesssim
6$\,GeV, $v_2(p_T)$ saturates and exhibits a remarkable scaling
between baryons and mesons~\cite{1st_scaling},
\begin{equation}
v_2(p_T)=n_q v_2^{(q)}(p_T/n_q) \ ,
\label{v2_pT_scaling}
\end{equation}
where $n_q$ denotes the number of valence quarks contained in hadron
$h$. Equation (\ref{v2_pT_scaling}), also known as
``constituent-quark number scaling" (CQNS), has been successfully
described in the framework of quark coalescence models, where quarks
recombine into hadrons at the transition from partonic to hadronic
matter~\cite{coalescencemodels}. In line with
Eq.~(\ref{v2_pT_scaling}), these models thus imply that the observed
hadron elliptic flow can be reduced to a universal quark elliptic
flow at hadronization. This assertion, however, requires several
simplifying assumptions. First, interactions in the subsequent
hadronic evolution (between hadronization at $T_c\simeq170$\,MeV and
kinetic freeze-out at $T_{\rm fo}\simeq 110$\,MeV) are neglected.
Furthermore, Eq.~(\ref{v2_pT_scaling}) can only be dervied in
instantaneous quark coalescence models where energy conservation is
violated and narrow hadron wave functions in momentum space are
utilized. In addition, the underlying quark distribution functions
are assumed to factorize in coordinate and momentum space, implying
that the elliptic flow of the system is implemented point-by-point
(i.e., locally), rather than through a space-dependent flow field
which is a consequence of a collectively expanding source. The
implementation of realistic space-momentum correlations into quark
coalescence models has indeed been proved
difficult~\cite{Pratt:2004zq}. Finally, quark coalescence models
typically do not comply with the principle of detailed balance,
implying that the proper equilibrium limit cannot be established.
This hampers attempts to extend the description into the low-$p_T$
regime where thermalization is believed to prevail. This has become
a more pressing issue once it was realized that CQNS can be
generalized to encompass both low and intermediate $p_T$ if $v_2$ is
plotted versus transverse kinetic energy, $\ket\equiv m_T-m$, where
$m_T=(p_T^2+m^2)^{1/2}$ is the hadron's transverse
mass~\cite{Phenix_KET_scaling2007,STAR_KET_scaling2008,Lacey2006,Ma2009,Steinberg_KET_scaling2008,Jia2007}.
In this representation, elliptic-flow data for all observed hadrons
fall onto one universal curve for $v_2/n_q$ versus
$\ket/n_q$~\cite{Phenix_KET_scaling2007,STAR_KET_scaling2008}.

Progress on the above issues has been recently reported by
reformulating the quark-coalescence process at hadronization in
terms of resonant $q\bar q\to M$ scattering ($M$:
meson)~\cite{Ravagli:2007xx,Cassing}. Implementing the underlying
cross section into a Boltzmann transport equation (e.g., in
Breit-Wigner approximation) not only guarantees energy-momentum
conservation but also satisfies detailed balance and thus leads to
the correct thermal-equilibrium limit. It has been
verified~\cite{Ravagli:2007xx} that the meson spectra resulting from
the Resonance Recombination Model (RRM) are compatible with CQNS,
Eq.~(\ref{v2_pT_scaling}), in the factorized (local) approximation
to $v_2(p_T)$. A further step has been taken in
Ref.~\cite{Ravagli:2008rt} where the RRM has been applied to more
realistic quark distribution functions as generated by relativistic
Langevin simulations within an expanding QGP
fireball~\cite{vanHees:2005wb} for semi-central Au-Au collisions at
RHIC. Since the Brownian-motion approximation underlying the
Langevin process is only reliable for relatively massive and/or
high-momentum particles, $m_T\gg T$, the results have been
restricted to charm and strange quarks. Under inclusion of the full
space-momentum correlations imprinted on the quark distributions by
the hydro-like flow fields of the expanding fireball, the elliptic
flow of $\phi$ and $J/\psi$ mesons was found to exhibit CQNS in
$\ket$ from 0 to $\sim$3\,GeV, which encompasses both low and
intermediate $p_T$, i.e., the quasi-thermal and kinetic regimes.
However, several important questions remain, e.g., the manifestation
of CQNS and $\ket$ scaling for bulk (light) particles in the
low-$p_T$ (thermal) regime, the robustness and generality of the
flow fields utilized in the fireball simulations or the role of
reinteractions in the hadronic phase. A thorough understanding of
these issues, in connection with a quantitative description of
hadron data, is hoped to ultimately enable the extraction of quark
distribution functions just prior to the conversion to hadronic
degrees of freedom, and thus to quantitatively establish the
presence of a collectively expanding partonic source in URHICs.

In the present paper we conduct investigations of scaling properties
of hadron spectra by focusing on the low-$p_T$ regime coupled with
the assumption of complete kinetic equilibration. We will utilize
the RRM to convert locally equilibrated quark distribution functions
into hadron spectra using a blast-wave type quark source with
realistic flow fields at the hadronization transition (including
space-momentum correlations characteristic of hydrodynamic
simulations). Contrary to earlier work~\cite{Ravagli:2008rt}, we do
not attempt to generate the flow fields from a dynamic evolution,
but rather extract them from empirical fits to hadron $p_T$-spectra
and elliptic flow. In fact, after verifying that resonance
recombination converts collective, locally equilibrated quark
distribution functions into equilibrium hadron distributions with
{\rm identical} (space-momentum dependent) collective properties,
one can adopt the source parametrization at the hadron level and
thus readily extend the description to baryons. For this part of the
investigation we focus on multi-strange hadrons (i.e., hadrons
containing at least 2 strange and/or anti-strange quarks) which are
believed to kinetically decouple close to the hadronization
transition. This notion is supported by empirical blast-wave fits to
RHIC and SPS data~\cite{RHIC2005,Xu2009,Olga_sequential}, as well as
by the putative absence of resonances that these hadrons could form
in interactions with bulk particles such as pions, kaons or nucleons
(hadronic resonances drive the collective expansion of the hadronic
phases in URHICs). We refer to hadrons decoupling close to $T_c$ as
``group-I" particles, which are thus the prime candidates to infer
properties of the quark phase just above the critical temperature.
On the other hand, the most prominent bulk particles ($\pi$, $K$,
$p$, etc.) are well-known to interact quasi-elastically via strong
hadronic resonances, such as $\rho$(770), $K^*$(892) and
$\Delta$(1232), which renders their interactions operative until a
kinetic freeze-out temperature, $T_{\rm fo}$, which is significantly
lower than the one of group-I particles, $T_{\rm fo}\simeq 110\,{\rm
MeV} < T_c \simeq 170\,{\rm MeV}$. We refer to hadrons decoupling at
$T_{\rm fo}$ as ``group-II" particles. Clearly, a description of
$\ket$ scaling in the low-$p_T$ regime would be incomplete (to say
the least) without $\pi$, $K$ and $p$. In an attempt to include
these particles, we adopt the simplest possible scenario of
introducing a second source for group-II particles by fitting their
spectra and $v_2$ with a space-dependent elliptic flow field at
$T_{\rm fo}$. We then evaluate in how far such a minimalistic (but
in our opinion well-motivated) sequential freeze-out can be
compatible with CQNS and $\ket$ scaling, and which mechanisms could
be responsible for them. In doing so we also revisit the question of
resonance feed-down contributions to the pion $v_2$ for which
differing results seem to exist in previous works.

Our article is organized as follows. In Sec.~\ref{sec_rrm} we
recapitulate basic features of the RRM, specifically its equilibrium
limit in the presence of anisotropic flow fields. In
Sec.~\ref{sec_group1} we quantitatively construct realistic flow
fields for semi-central Au-Au collisions at RHIC using group-I
particle freeze-out close to the hadronization transition. In
Sec.~\ref{sec_group2} we reiterate this procedure for group-II
particles at kinetic freeze-out. In Sec.~\ref{sec_KET-scal} we
examine our results for group-I and -II particles with respect to
CQNS and $\ket$ scaling. The effects of feed-down on the inclusive
pion $v_2$ are revisited in App.~\ref{app_feeddown}. We summarize
and conclude in Sec.~\ref{sec_concl}.

\section{Resonance Recombination Model and Equilibrium Limit}
\label{sec_rrm}
Early quark coalescence models which employed instantaneous
approximations in the recombination of
quarks~\cite{coalescencemodels,more_coalescence} were quite
successful in providing an explanation for two rather unexpected
phenomena in hadron spectra at intermediate $p_T$ in Au-Au
collisions at RHIC. \footnote{Throughout this manuscript, we will
use $p_T$ ($p_t$) to denote hadron (quark) transverse momenta.}
Specifically, these were the enhancement of baryon-to-meson ratios
($B/M$ = $p/\pi$, $\Lambda/K$) over their values measured in $p$-$p$
collisions, as well as CQNS of $v_2(p_T)$. While an enhanced $B/M$
ratio is also a natural outcome of hydrodynamic flow, the latter is
not expected to generate enough yield to dominate hadron production
at $p_T>2-3$\,GeV. The instantaneous projection of parton states
onto hadron states conserves 3-momentum by construction but energy
conservation is violated due to the sudden
approximation~\cite{Pratt:2004zq}. This approximation is believed to
be viable at intermediate $p_T$ where corrections are expected to be
of order ${\cal O}(Q/p_T)$ with $Q=m-m_q-m_{\bar{q}}$ ($m$: meson
mass, $m_{\bar{q},q}$: anti-/quark mass), albeit with unknown
coefficient~\cite{Ravagli:2007xx}. However, this poses a significant
problem at low $p_T$ and/or for hadrons (resonances) with large
binding energy ($Q$-value).

In Ref.~\cite{Ravagli:2007xx} quark coalescence in an interacting medium
in the vicinity of the hadronization transition was reinterpreted as a
process akin to the formation of resonances,
$q+\overline{q}\rightleftharpoons M$, and implemented via a Boltzmann
equation
\begin{equation}
p^{\mu}\partial_{\mu}f_M(t,\vec x,\vec p)=-m\Gamma f_M(t,\vec x,\vec
p)+p^0\beta(\vec x, \vec p),
\label{Boltzmann}
\end{equation}
where $f_M(t,\vec x, \vec p)$ denotes the phase space density of
the meson, and the gain term is given by
\begin{eqnarray}
\beta(\vec x,\vec p)=\int\frac{d^3p_1d^3p_2}{(2\pi)^6}f_q(\vec
x,\vec p)f_{\bar q}(\vec x,\vec p)\nonumber \\
\times\sigma(s)v_{\rm rel}(\vec p_1,\vec p_2)\delta^3(\vec p -\vec p_1
-\vec p_2) \ .
\label{gainterm}
\end{eqnarray}
The use of an explicit (resonant) cross section automatically
satisfies energy-momentum conservation. For simplicity it has been
modeled by a relativistic Breit-Wigner form,
\begin{equation}
\sigma(s)=g_{\sigma}\frac{4\pi}{k^2}\frac{(\Gamma
m)^2}{(s-m^2)+(\Gamma m)^2} , \
\label{Breit-Wigner}
\end{equation}
where the same reaction rate $\Gamma$ is used as in the loss term,
the first term on the right-hand-side ($rhs$) of
Eq.~(\ref{Boltzmann}), and $g_{\sigma}$ is the degeneracy factor.
This guarantees detailed balance, which is essential for recovering
thermodynamic equilibrium in the long-time limit, $\Delta \tau \gg
1/\Gamma$. If hadronization is rapid enough to produce hadrons in
equilibrium, as it seems to be the case for the bulk of hadrons in
URHICs, this limit is applicable and the resulting hadron-momentum
distribution is given by
\begin{equation}
f_M^{eq}(\vec p)=\frac{E_M(\vec p)}{m\Gamma}\int d^3x\beta(\vec
x,\vec p)
\label{rrm-pt}
\end{equation}
For more details, we refer the reader to Ref.~\cite{Ravagli:2007xx}.
It is well known that the unique equilibrium solution of a Boltzmann
transport equation is a Boltzmann distribution if and only if energy
conservation as well as detailed balance are satisfied. The RRM
complies with these requirements. In the remainder of this section
we will verify numerically that Eq.~(\ref{rrm-pt}) recovers the
thermal Boltzmann distribution for mesons in the presence of a
spatially dependent anisotropic flow field at the quark level (in
previous work~\cite{Ravagli:2007xx,Ravagli:2008rt}, this has only
been exemplified for $p_T$ spectra, not for $v_2$).

\begin{figure}[!t]
\includegraphics[width=\columnwidth
]{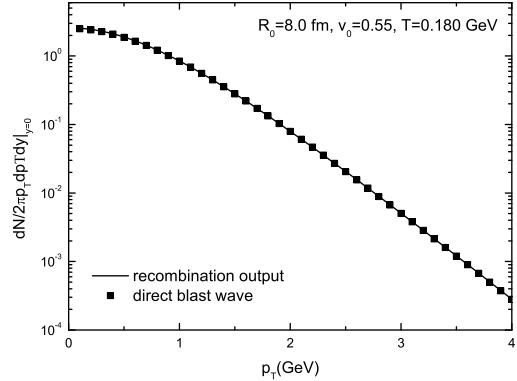}
\caption{(Color online) Comparison of $\phi$-meson $p_T$ spectra
computed from (i) resonance recombination of strange quarks using
blast-wave flow fields for strange quarks (rectangles), and
(ii) a direct application of an equilibrium blast-wave distribution
(solid line) with  the same flow field as for quarks (solid line).
The hadronization temperature has been chosen at $T_c = 180$\,MeV and
the radial flow at the surface as $v_0 = 0.55$.}
\label{fig_rrm-pT}
\end{figure}

As a first test we study $\phi$ mesons in an azimuthally symmetric
and longitudinally boost-invariant fireball. The phase
space-densities of the input strange and anti-strange quarks
($m_s=0.45$\,GeV) are parameterized through a blast-wave with radial
flow velocity $\vec v(\vec r)=v_0\cdot\vec r/R_0$. We compare the
result for $\phi$ mesons ($m_\phi = 1.02$ GeV, $\Gamma_\phi =
0.05$\,GeV) calculated from resonance recombination of $s$ and $\bar
s$, Eq.~(\ref{rrm-pt}), with the direct blast-wave expression for
the $\phi$ meson with identical flow field and temperature as for
the quark input. Fig.~\ref{fig_rrm-pT} shows that both ways of
computing the $\phi$ spectra are in excellent agreement. We have
reconfirmed that the results are insensitive to variations of the
resonance width $\Gamma$ when ensuring $\Gamma \lesssim Q$ with
$Q=m_{\phi}-(m_s+m_{\bar{s}})$~\cite{Ravagli:2007xx}.

\begin{figure}[!t]
\includegraphics[width=\columnwidth
]{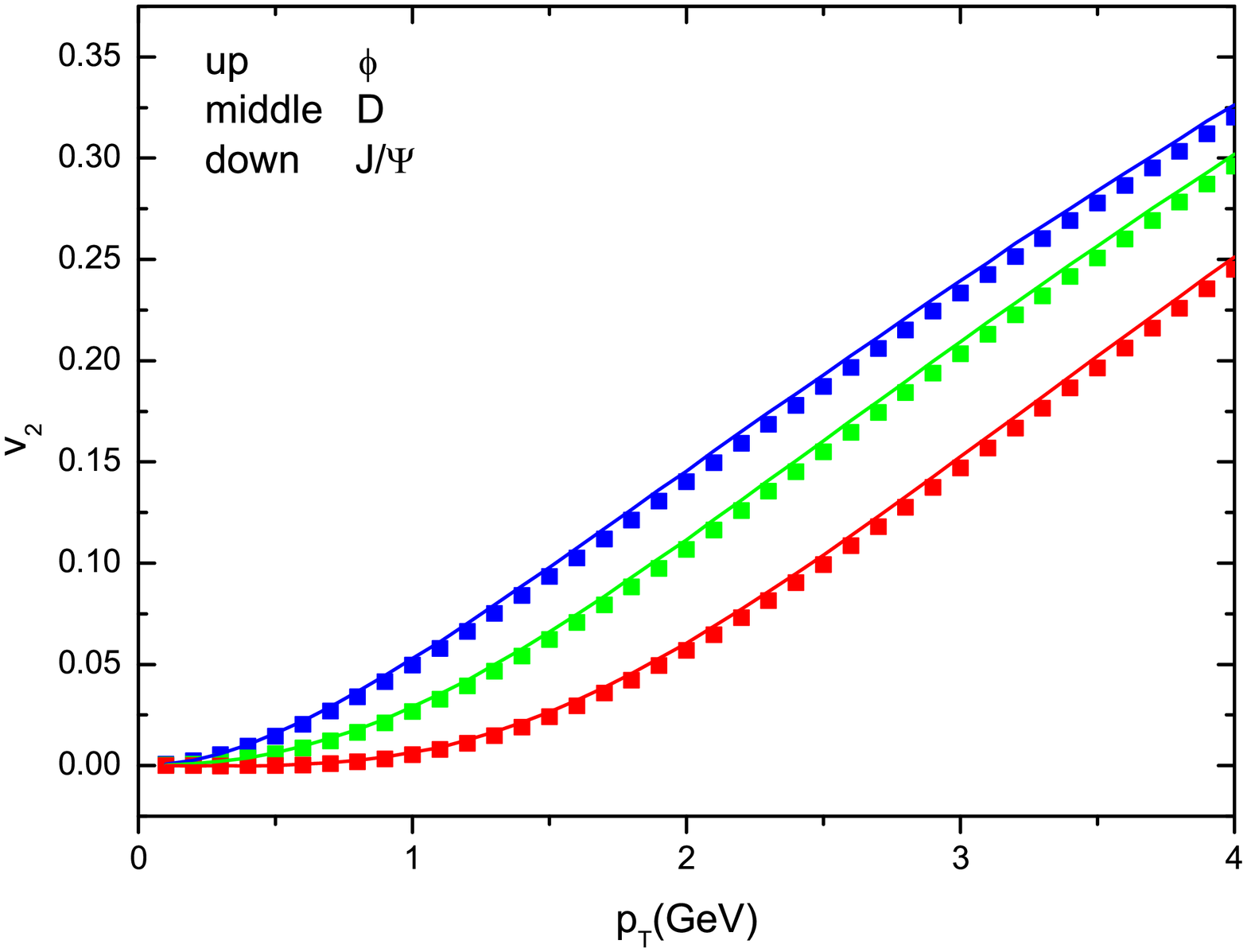}
\caption{(Color online) Comparison of elliptic flow for $\phi$, $D$ and
$J/\psi$ mesons (top to bottom) computed from:
(i) resonance recombination of strange quarks (symbols), and (ii) a direct
application of the blast-wave expression at the meson level (lines).
Both quark and meson distributions are based on the same anisotropic
flow field constructed from the RL parameterization with $T_c = 180$\,MeV,
$R_{\rm x}=7.5$\,fm, $R_{\rm y}=8.5$\,fm, $\rho_0=0.66$ and
$\rho_2=0.07$.}
\label{fig_rrm-v2}
\end{figure}

To account for anisotropic flow fields which are suitable for
generating configurations reminiscent of hydrodynamic simulations
for fireballs in non-central (azimuthally asymmetric) nuclear
collisions, we adopt in this work the parameterization introduced by
Reti\`ere and Lisa (RL)~\cite{Lisa2004}. It assumes longitudinal
boost-invariance~\cite{Bjorken1984} and parameterizes the transverse
flow rapidity as a function of radius $r$ and spatial azimuthal
angle $\phi_s$ as
\begin{equation}
\rho(r,\phi_s)=\widetilde{r}[\rho_0+\rho_2 \cos (2\phi_b)] \ ,
\label{transverse_rapidity}
\end{equation}
where $\rho_0$ represents the average radial flow rapidity and
$\rho_2$ the anisotropy of transverse flow. The angle $\phi_b$ of
the flow vector is generally tilted from the position angle $\phi_s$
(both angles are measured with respect to the reaction plane) which
can be reflected by the relation
\begin{equation}\label{flow_angle}
  \tan \phi_b=\left(\frac{R_{\rm x}}{R_{\rm y}}\right)^2 \tan \phi_s \ .
\end{equation}
This ansatz is motivated by the picture that the transverse boost is
locally perpendicular to elliptical subshells~\cite{Lisa2004}.
The transverse flow rapidity, $\rho(r,\phi_s)$, is assumed to
increase linearly with the normalized "elliptic radius", defined as
\begin{equation}
\widetilde{r}=\sqrt{r^2~\cos^2\phi_s/R_{\rm
x}^2+r^2\sin^2 \phi_s/R_{\rm y}^2} \ .
\label{normalized_radius}
\end{equation}
We emphasize that this flow field incorporates space-momentum
correlations that are considered a much more realistic
representation of a collectively expanding source than the
factorized ansatz of a ``local" $v_2$ (independent of position) as
often adopted in quark coalescence models calculations.

Using the RL parameterization the differential momentum spectrum for
a particle $i$ directly emitted from the source takes the
form~\cite{Lisa2004,Heinz20032004,Heinz1993} (feed-down
contributions are evaluated in App.~\ref{app_feeddown})
\begin{eqnarray}
\frac{dN_i}{p_Tdp_Td\phi_pdy}=\frac{2g_i}{(2\pi)^3} \ \tau_f \  m_T
\
  {\rm e}^{\mu_i/T_f} \qquad \qquad \qquad
\nonumber\\
 \times \int rdr \int d\phi_s \ K_1(m_T, T, \beta_T) \
{\rm e}^{\alpha_T\cos (\phi_p-\phi_b)} \ ,
\label{pTspectrum}
\end{eqnarray}
where $T_f$ is the freeze-out temperature at constant longitudinal
proper time $\tau_f$, $\mu_i$ the chemical potential of particle
$i$, $g_i$ the spin-isospin degeneracy factor, $K_1$ a modified
Bessel function and $\alpha_T=p_T/T_f \sinh \rho(r,\phi_s)$,
$\beta_T=m_T/T_f \cosh \rho(r,\phi_s)$.
Equation~(\ref{pTspectrum}) has been written in Boltzmann approximation
which works well for sufficiently heavy particles; we will, however,
use Bose distributions for pions. From Eq.~(\ref{pTspectrum}) we can
calculate the elliptic flow of particle $i$ as
\begin{equation}\label{v2}
v_2^{i}(p_T)=\frac{\int_0^{2\pi}d\phi_p \cos
(2\phi_p)\frac{dN_{i}}{p_Tdp_Td\phi_p dy}}{\int_0^{2\pi}d\phi_p
\frac{dN_{i}}{p_Tdp_Td\phi_p dy}} \, .
\end{equation}

With the RL parameterization for an anisotropic flow field we are
now in position to compare the elliptic flow generated for coalesced
mesons in the RRM for locally equilibrated quark distributions,
Eq.~(\ref{rrm-pt}) with the one directly obtained from applying the
blast-wave expression (\ref{pTspectrum}) at the meson level. We fix
the parameters as quoted in the caption of Fig.~\ref{fig_rrm-v2},
noting that at this point our aim is not to fit experimental data
but to scrutinize generic properties of the RRM. We emphasize again
the nontrivial spatial dependence of the flow field figuring into
the quark distributions in Eq.~(\ref{rrm-pt}). To extend the scope
of our comparison we consider, in addition to $\phi$ mesons, also
$D$ ($m_D = 1.9$\,GeV, $\Gamma_D = 0.1$\,GeV) and $J/\psi$ mesons
($m_{J/\psi} = 3.1$\,GeV, $\Gamma_{J/\psi} = 0.1$\,GeV), as bound
states of charm ($m_c = 1.5$\,GeV) and light quarks ($m_{u,d} =
0.3$,\,GeV) and of charm and anticharm quarks, respectively.
Fig.~\ref{fig_rrm-v2} illustrates the comparison of the
meson-$v_2(p_T)$ for RRM and their direct blast-wave counterparts,
showing again excellent agreement within numerical accuracy. It is
quite remarkable that RRM generates negative $v_2$ at low $p_T$ for
the $J/\psi$ starting from strictly positive $v_2(p_t)$ for the
underlying charm-quark distributions. A possibly negative $v_2$ is a
well-known mass effect caused by the depletion of the heavy-particle
yield at low momenta in the presence of sufficiently large
collective flow~\cite{Heinz2001}. However, we are not aware of any
quark coalescence model calculation that has reproduced this result.
It demonstrates that the meson distributions resulting from RRM have
reached local equilibrium and follow the collective flow of the
source. We have checked that the negative $v_2$ is rather robust
against variations of blast-wave parameters. By increasing $\rho_0$
or $\rho_2$ the negative value can be amplified.

Combining the results from Figs.~\ref{fig_rrm-pT} and
\ref{fig_rrm-v2}, we conclude that hadron phase-space distributions
obtained from resonance recombination precisely reflect the
collective properties of a source in local equilibrium, explicitly
documenting the fact that the RRM achieves this through energy
conservation and detailed balance. These insights also imply that
the intermediate-$p_T$ regime, where $v_2$ saturates and shows an
explicit dependence on quark number, can not be in full equilibrium.

\section{Recombination and Quark Distributions at Hadronization}
\label{sec_group1}
Based on the verification that resonance recombination maps the
collective local-equilibrium properties of the quark source into
hadron spectra, we can reverse the strategy and use this as a tool
to extract quark distributions by a quantitative study of
experimental data on hadrons which arise from quark coalescence. The
generality of this argument even allows to extend the analysis to
baryons without having to explicitly calculate their spectra in RRM
(which would be more involved than for mesons as it requires to
recombine 3 quarks). As alluded to in the introduction, this
procedure can only give access to quark distributions if the
subsequent hadronic phase exerts a negligible distortion of the
hadron spectra right after their formation around $T_c$. We label
such hadrons as belonging to ``group I", and identify multi-strange
hadrons ($\phi$, $\Xi$, $\Omega$) as the prime candidates currently
available from experiment. These hadrons have no well-established
resonance excitations on the most abundant bulk particles ($\pi$,
$K$ and $N$) in the hadronic medium, while elastic $t$-channel
exchange processes are suppressed by the OZI rule. One can therefore
expect that multi-strange hadrons do not suffer significant
rescattering in the hadronic phase, which is indeed supported by
blast-wave fits to experimental data favoring kinetic decoupling not
far from $T_c$. The flow fields extracted for group-I hadrons
therefore reflect the collective properties of the quark phase.

\begin{table}[tb]
\begin{tabular}{|c||c|c|c|c|c|}
  \hline
  Group & $T_f$ & $\rho_0$ & $\rho_2$ & $R_{\mathrm{x}}$ & $R_{\mathrm{y}}$ \\
  \hline\hline
  I & 180 & 0.75 & 0.072 & 7.5 & 8.9 \\
  \hline
  II & 110 & 0.93 & 0.055 & 11.0 & 12.4 \\
  \hline
\end{tabular}
\caption{Parameter values for kinetic freeze-out temperature, $T_f$
in MeV,
  radial-flow rapidity $\rho_0$, rapidity-asymmetry, $\rho_2$, and elliptic
  radii, $R_{\mathrm{x}}$ and $R_{\mathrm{y}}$ in fm, for group-I and -II
  hadrons using the RL blast-wave expression for mid-central Au-Au
  collisions at RHIC.}
\label{tab:RLfit}
\end{table}

\begin{figure}[!t]
\hspace{4mm}
\includegraphics[width=\columnwidth
]{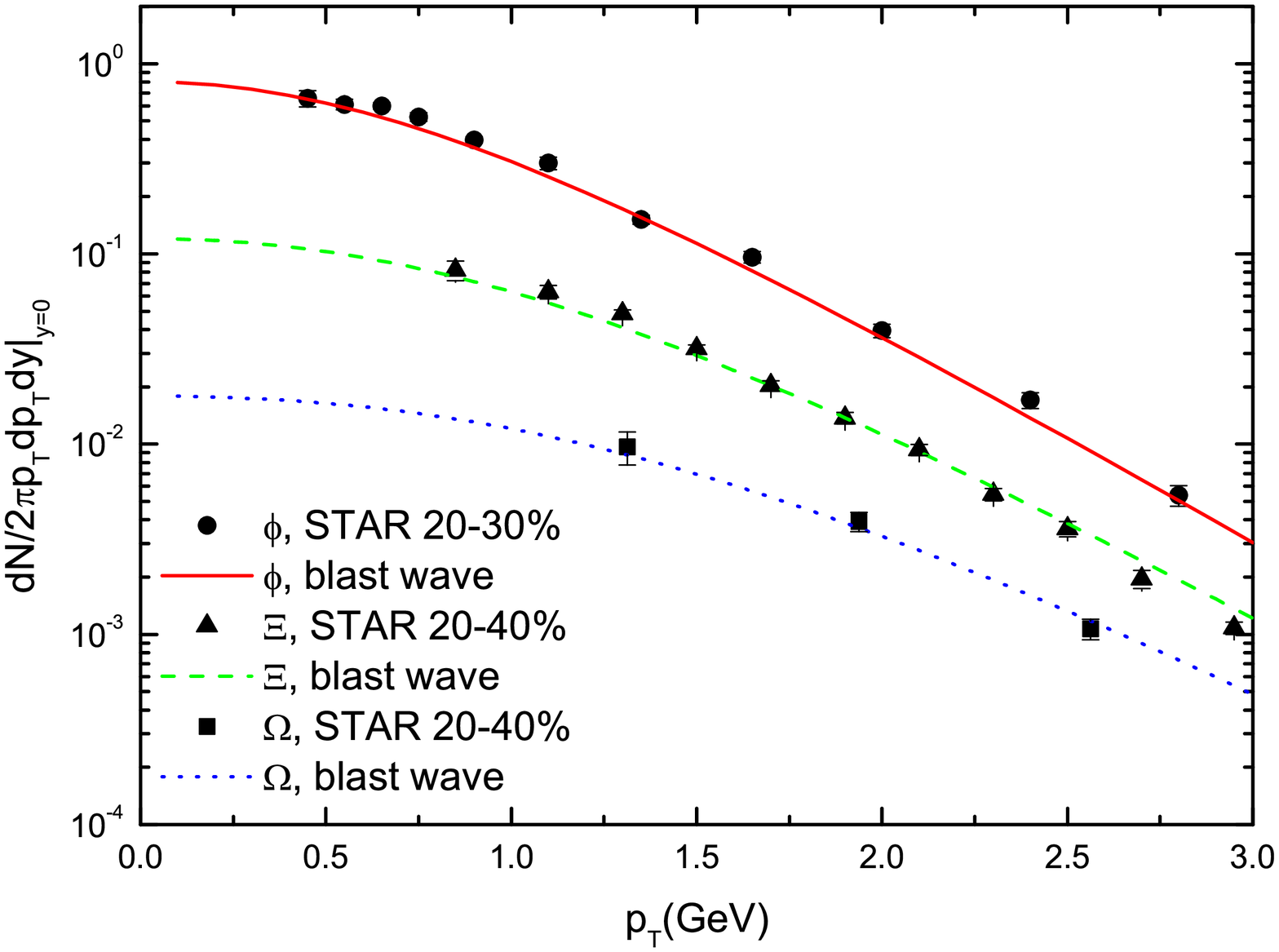}
\includegraphics[width=\columnwidth
]{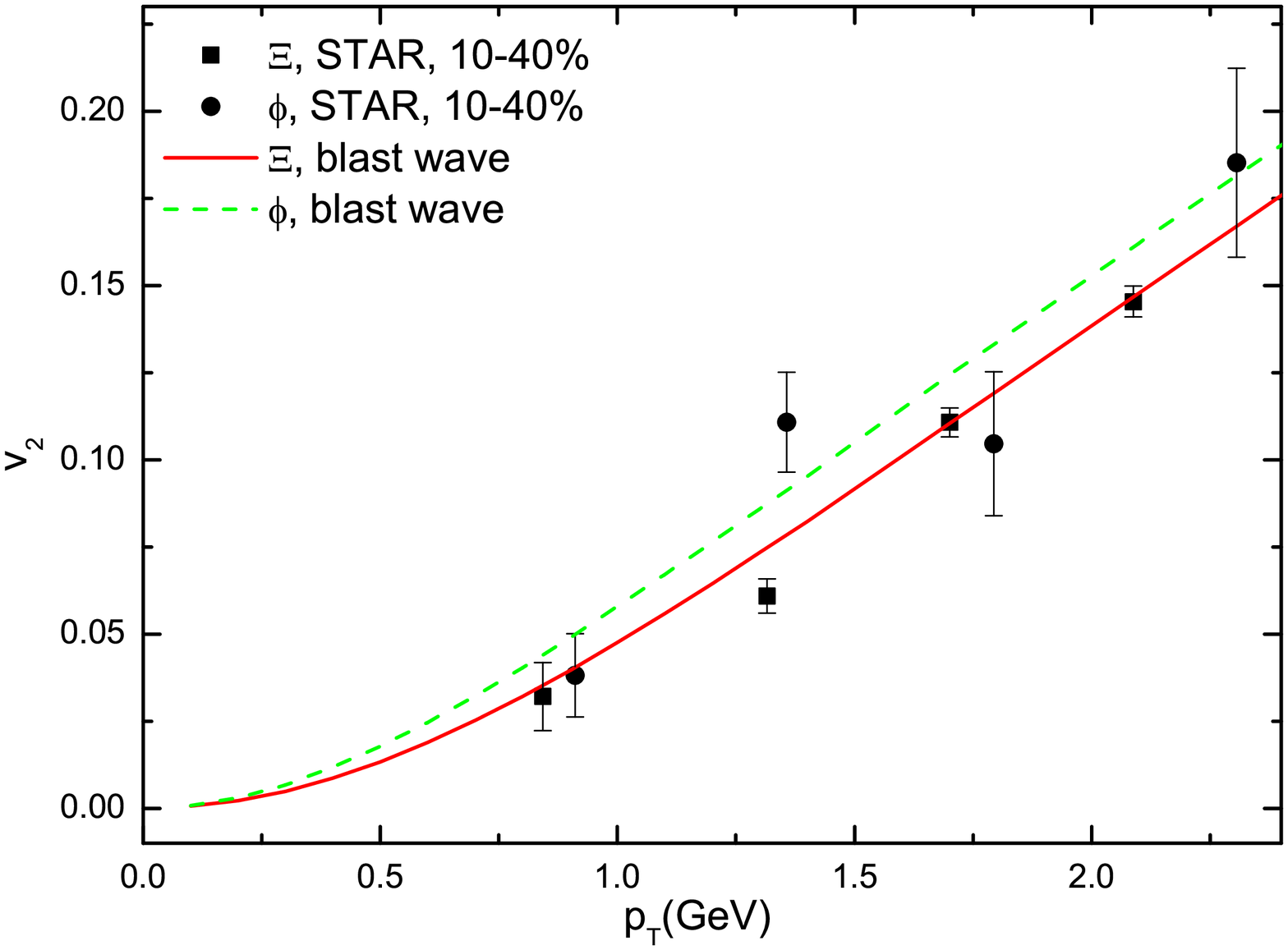}
\includegraphics[width=\columnwidth
]{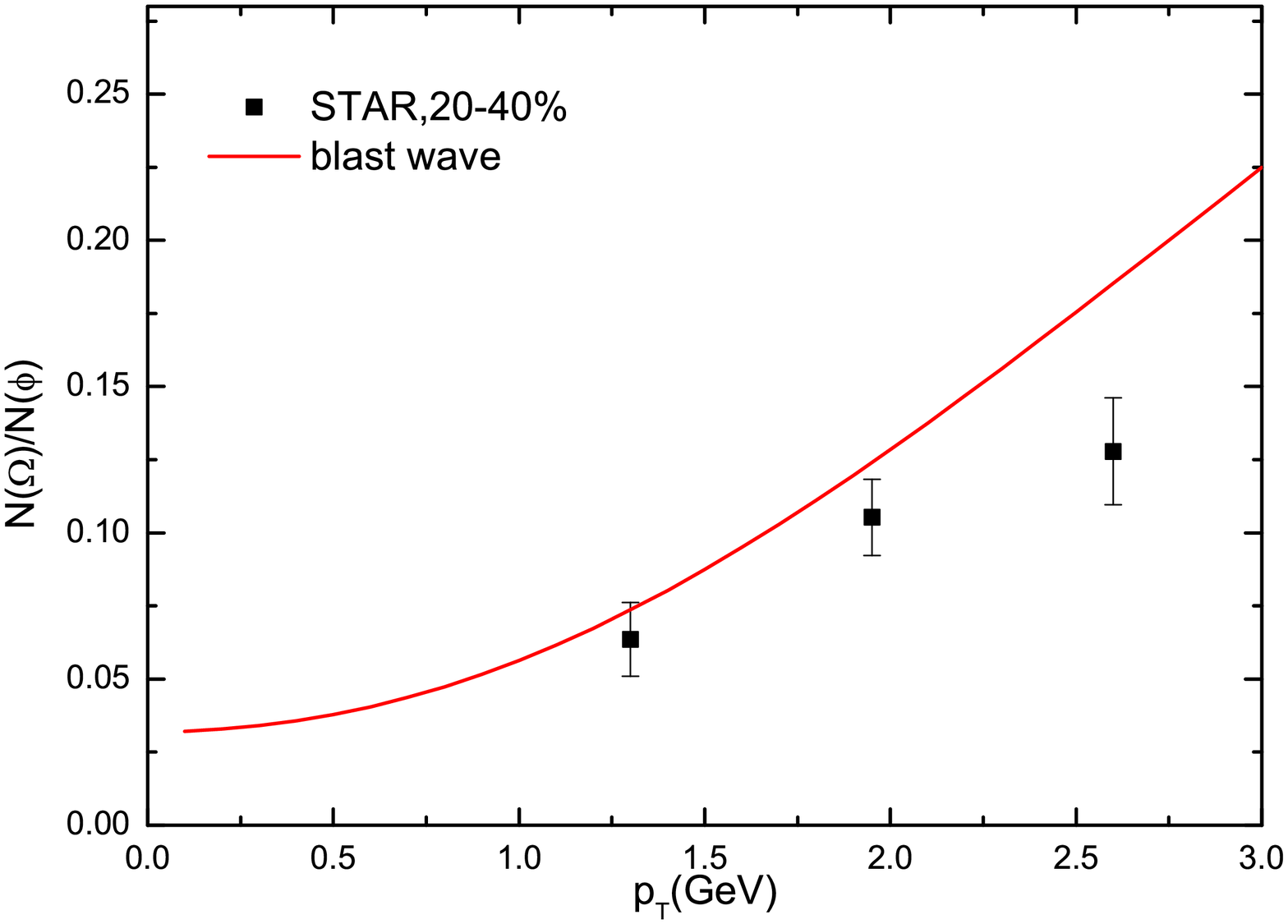} \caption{(Color online) Blast-wave fits to STAR
data~\cite{STAR_multistrange_pT,STAR_phi,STAR_phi_new,STAR_KET_scaling2008}
in semi-cental Au-Au($\sqrt{s_{NN}}$=200\,GeV) collisions for
transverse-momentum spectra (upper panel) and elliptic flow (middle
panel) of $\phi$, $\Xi^+$ and $(\Omega+\bar\Omega)/2$ using the RL
ansatz with parameters specified in Tab.~\ref{tab:RLfit}.  The lower
panel shows the resulting $\Omega$-to-$\phi$ ratio on a linear
scale, compared to STAR data~\cite{STAR_phi}.} \label{fig_g1-fit}
\end{figure}
We first determine the flow parameters in the RL parameterization by
performing a blast-wave fit to the transverse-momentum spectra and
elliptic flow of  group-I hadrons to experimental data from STAR in
mid-central (20-40\%) Au-Au collisions. For definiteness (and easier
comparison to the RRM calculations in Ref.~\cite{Ravagli:2008rt}),
we fix the temperature to $T_c=180$\,MeV; we do not utilize it as a
parameter but obtain fits of satisfactory quality with this value.
The transverse-flow rapidity, $\rho_0$, is mostly governed by fits
to the $p_T$-spectra of $\phi$, $\Xi$ and $\Omega$ hadrons, while
the flow-asymmetry parameter, $\rho_2$, and the fireball's spatial
eccentricity, $R_{\mathrm y}/R_{\rm x}$, are driven by fitting the
elliptic flow of $\Xi$ and $\phi$. The resulting RL blast-wave
curves are shown with data in Fig.~\ref{fig_g1-fit}, and the central
values of the parameters are collected in Tab.~\ref{tab:RLfit}. As
to be expected, the slopes of the spectra can be fitted well. The
absolute normalizations are also reproduced reasonably as well,
which is confirmed by the $\Omega$-to-$\phi$ ratio shown on a linear
scale in the lower panel of Fig.~\ref{fig_g1-fit}. Degeneracy
factors for spin and isospin are taken into account, but baryon
chemical potential and the strangeness suppression factor,
$\gamma_s$, are neglected in these fits (they contribute at the 10\%
level which can easily be absorbed by fine-tuning the freeze-out
temperature). Note that the $\Omega$-to-$\phi$ ratio is quite
different from linear for $p_T$ up to $\sim$1.5\,GeV, and that this
ratio does not go to zero for $p_T\to 0$ as predicted by quark
coalescence models discussed in Ref.~\cite{Hwa_Yang,STAR_phi}. This
reiterates once more the importance of energy conservation and the
thermal-equilibrium limit if on attempts to describe hadron spectra
through quark coalescence in the low-$p_T$ regime.

\begin{figure}[!t]
\hspace{4mm}
\includegraphics[width=\columnwidth
]{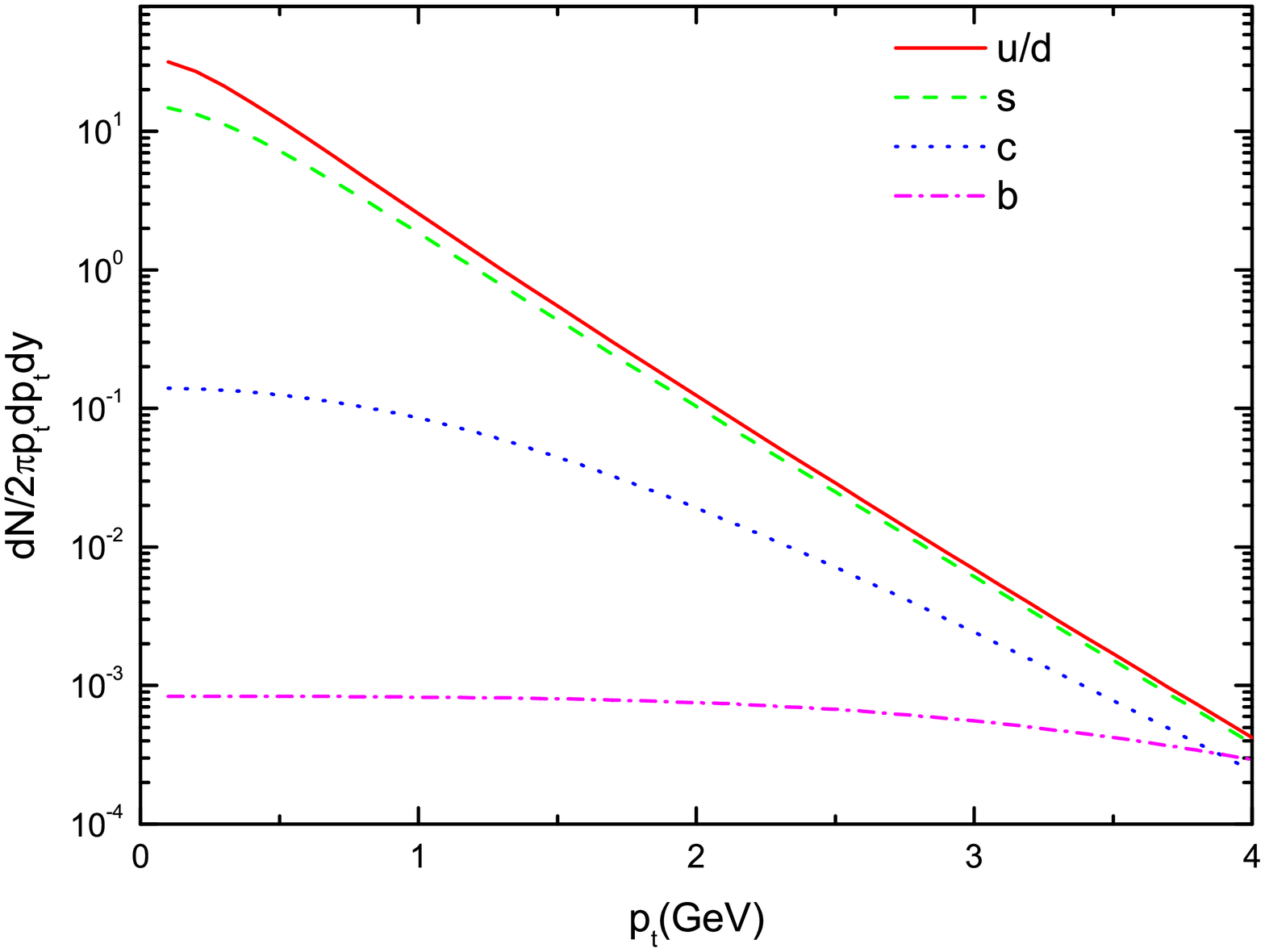}
\includegraphics[width=\columnwidth
]{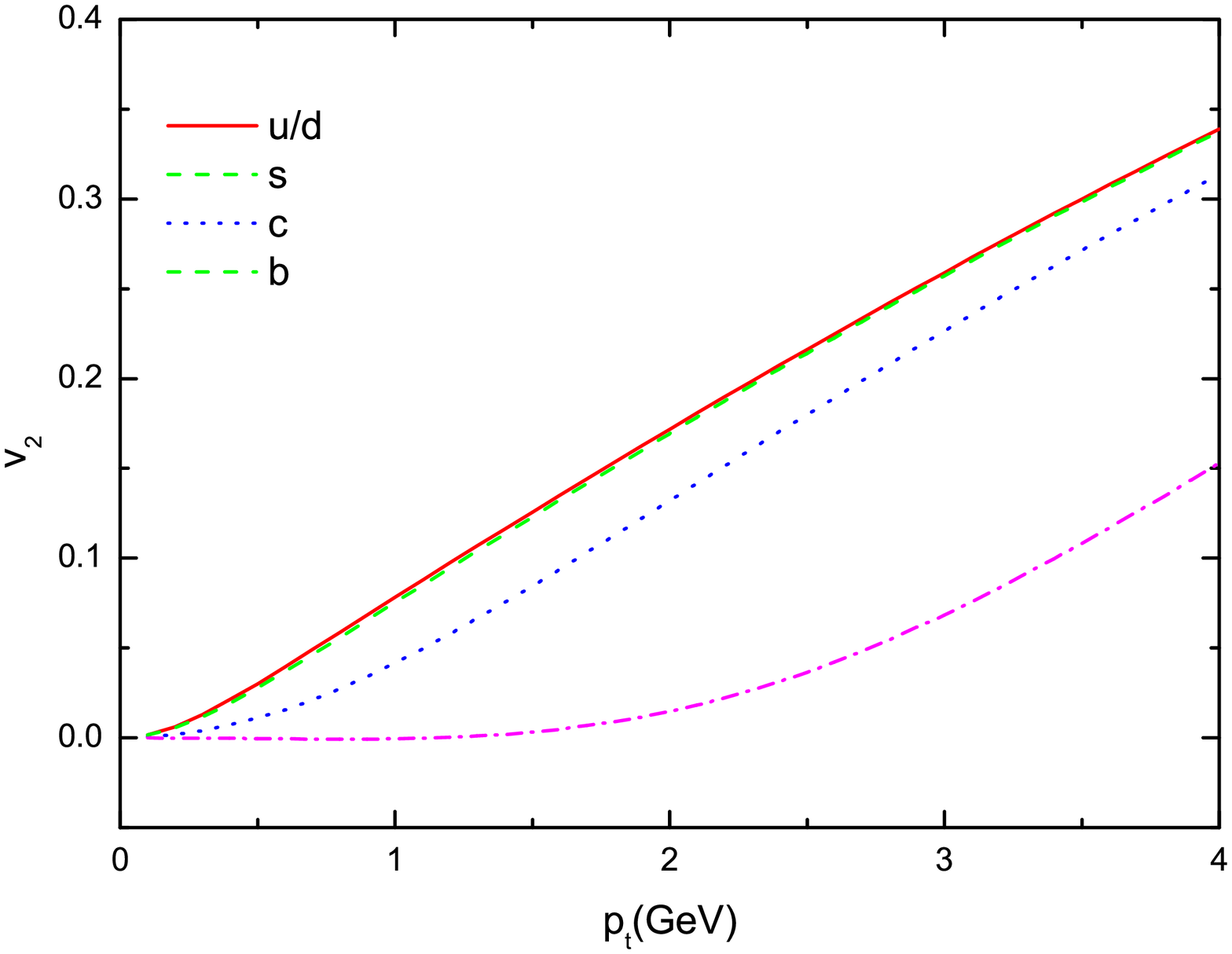}
\caption{(Color online)
Upper panel: Equilibrated transverse momentum distributions of quarks
on a hypersurface with temperature just above $T_c$ for a mid-central
fireball at RHIC energies as predicted by the RRM model from group I data.
Bottom panel: The same for the $v_2$ of the equilibrated quarks.}
\label{strange_pT}
\end{figure}

We are now in a position to extract the quark distribution functions
at hadronization by assuming that: (i) the flow field for group-I
hadrons is indeed the flow field for the hadronic phase in
equilibrium just below $T_c$ (which is further supported by the fact
that it is possible to use a freeze-out temperature of 180\,MeV),
and (ii) group-I hadron formation occurs via coalescence from a
partonic phase (which is supported by the general CQNS
characteristics of RHIC data). In Fig.~\ref{strange_pT} we not only
include the extracted strange- and light-quark spectra and elliptic
flow (which are directly involved in the RRM-based fits) but also
those implied for the heavier charm and bottom quarks. Equilibration
of the latter 2 quark flavors is currently an open issue, but this
may rather be a question of how far up in $p_t$ it applies.
Qualitatively, the same limitation applies to light and strange
quarks, which presumably enter the kinetic regime at $p_t\simeq
1~\rm GeV$, where the data for the quark-number scaled $v_2$ level
off at a value of about 7-8\%. Once available, $D$- and $B$-meson
data will allow for similar estimates, and we provide here the
underlying heavy-quark spectra as a reference for such an analysis.

An extraction of strange- and light-quark spectra in a similar
spirit, i.e., from data of what we refer to as group-I hadrons, has
been attempted by in Ref.~\cite{HZHuang} based on a 1-dimensional
instantaneous quark coalescence model. We reiterate here that such
an extraction cannot be reliably applied at low $p_T$ due to the
simplifying assumptions inherent to such models, but it would
certainly be interesting to make a comparison between both
approaches.

\section{Sequential Freeze-Out and Bulk Particles}
\label{sec_group2}
In URHICs the most abundant bulk particles, like pions, kaons and
protons (which we refer to as group-II particles), undergo
significant re-scattering in the hadronic phase. This is borne out
of experimental blast-wave fits which require kinetic-freeze-out
temperatures (e.g., $T_{\rm fo}\simeq100$\,MeV for central Au-Au
collisions at RHIC~\cite{RHIC2005}) which are significantly lower
than the chemical freeze-out temperatures ($T_{\rm ch}\simeq
160-180$\,MeV, extracted from observed ratios of different hadron
species). Note that the difference in the kinetic and chemical
freeze-out temperatures translates into a factor 5-10 different
energy densities~\cite{Rapp:2002fc}. Theoretically, the sequential
freeze-out systematics can be readily understood from large resonant
cross sections operative for group-II particles in the hadronic
phase, e.g., $\pi\pi\rightarrow\rho\rightarrow\pi\pi$, $\pi
K\rightarrow K^*\rightarrow \pi K$ or $\pi
N\rightarrow\Delta\rightarrow\pi N$. The values of these elastic
cross sections reach up to 100-200\,mb and are thus 1-2 orders of
magnitude larger than chemistry-changing inelastic ones (e.g.,
$\pi\pi\to KK$). Further direct evidence for an extended duration of
hadronic reinteractions follows from the well-established low-mass
dilepton enhancement in URHICs whose magnitude and spectral shape
requires several generations of $\rho$ mesons to form with a largely
broadened spectral shape due to a strong coupling to the
medium~\cite{Rapp:2009yu}.

In order to include group-II particles in an analysis of the thermal
regime of anisotropic flow fields there is thus no choice but to
introduce an additional source parameterization at a much lower
temperature than for group-I particles. While this generally
detaches group-II particles from coalescence mechanisms at
hadronization, we remark that their production through resonance
recombination in equilibrium at $T_c$ by no means contradicts but
rather facilitates local equilibrium until a much lower bulk
freeze-out. Group-II hadrons at low $p_T$ are therefore not suitable
to infer direct statements about quark recombination. Our aim rather
is to test whether the minimal assumption of a sequential freeze-out
is sufficient to produce the general scaling properties observed in
the data, and, if so, what the key features are and how sensitive
this picture is to variations of the freeze-out configuration. We
note in passing that the situation is presumably quite different at
intermediate $p_T$ where hadronic re-scattering is insufficient to
maintain equilibration and one is rather dealing with the
(transition to) a kinetic regime.

\begin{figure}[!t]
\hspace{4mm}
\includegraphics[width=\columnwidth
]{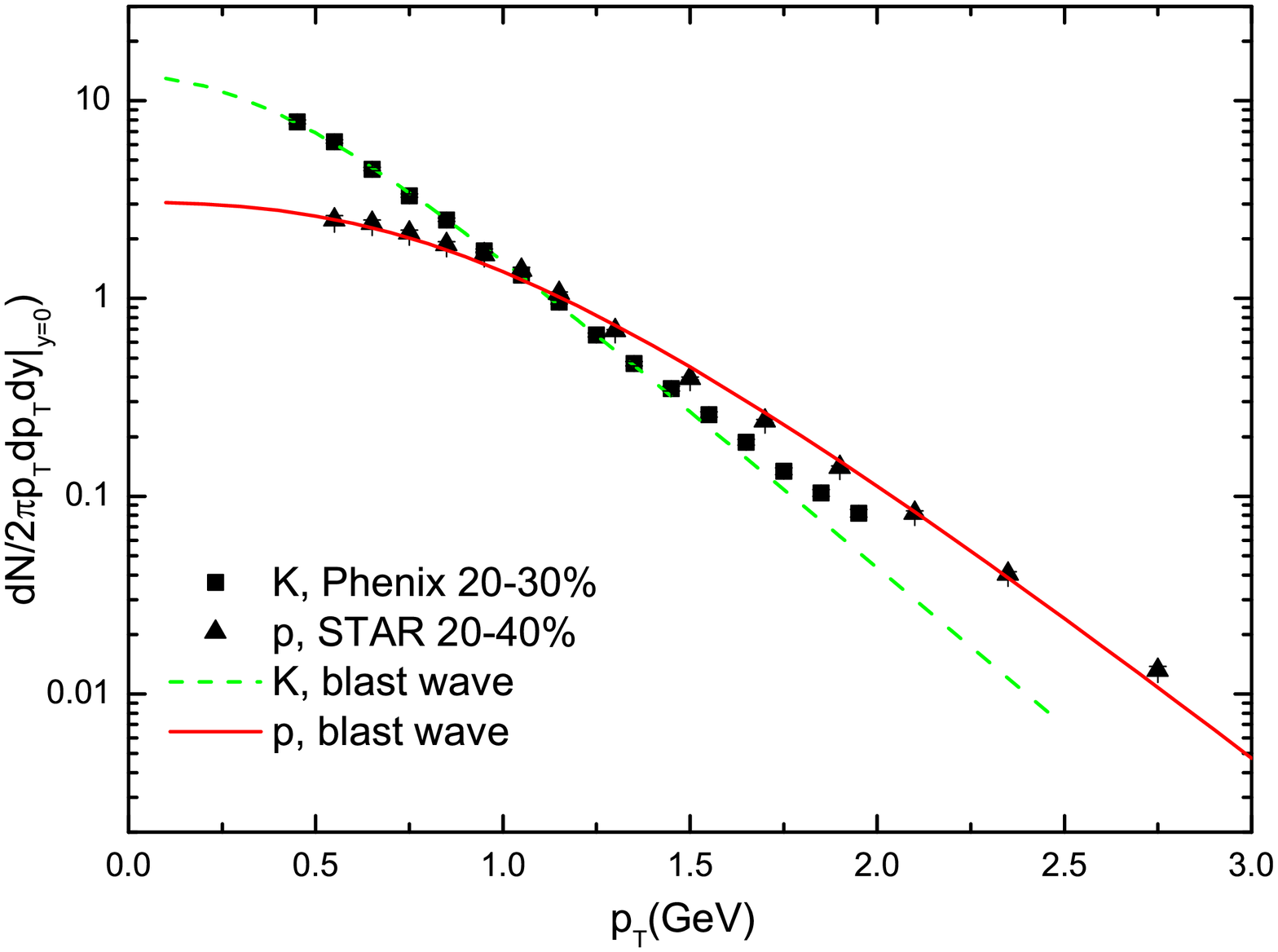}
\includegraphics[width=\columnwidth
]{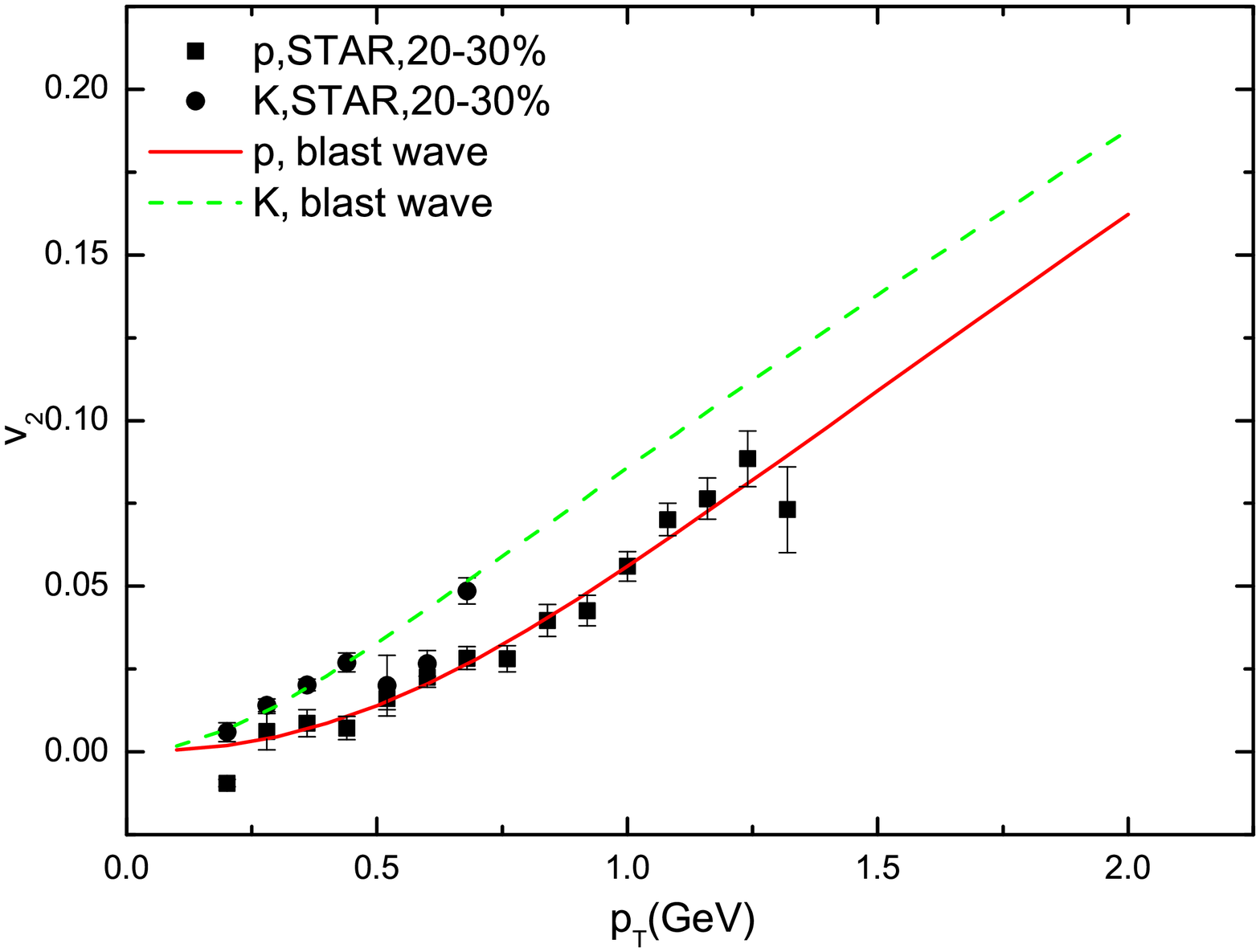} \caption{(Color online) Blast-wave fits using the RL
fireball parameterization to proton and kaon $p_T$-spectra (upper
panel) and elliptic flow (lower panel) in mid-central Au-Au
collisions at RHIC. The flow-field source parameters are listed in
Tab.~\ref{tab:RLfit}. Experimental data are taken from
Refs.~\cite{RHICv2,STAR_proton,kaon_pT}.} \label{group2_fit}
\end{figure}
Toward this end we again employ the RL blast-wave ansatz to
parameterize an anisotropic flow field which results in a good fit
to the $p_T$ spectra and elliptic flows of group-II particles. As in
the group-I case, we decide to fix the freeze-out temperature at a
value compatible with earlier blast-wave extractions, at $T_{\rm
fo}=110$\,MeV. Then we determine $\rho_0$, $\rho_2$ and the
eccentricity of the fireball from fits to the measured spectra and
$v_2$ of protons and kaons for the same centrality selection of
Au-Au collisions. The results are displayed in
Fig.~\ref{group2_fit}, corresponding to the parameter values listed
in Tab.~\ref{tab:RLfit}. At kinetic freeze-out, effective chemical
potentials for stable particles (pion, kaon, nucleon) are required
to correctly predict the absolute normalization of their
yields~\cite{Rapp:2002fc}. Neglecting chemical potentials therefore
implies that we can only fit the shape of the spectra which is
sufficient for our purposes here. In Boltzmann approximation, the
chemical potential appears as a constant overall fugacity factor
which does not affect the $v_2$ either. The pions are treated with
Bose statistics which entails an appreciable (moderate) enhancement
of the $p_T$ spectra ($v_2$) at $p_T\lesssim 0.5$\,GeV. At the same
time, feed-down contributions are also significant in this regime. A
more detailed discussion of the latter is given in
App.~\ref{app_feeddown}.

\begin{figure}[!t]
\hspace{4mm}
\includegraphics[width=\columnwidth
]{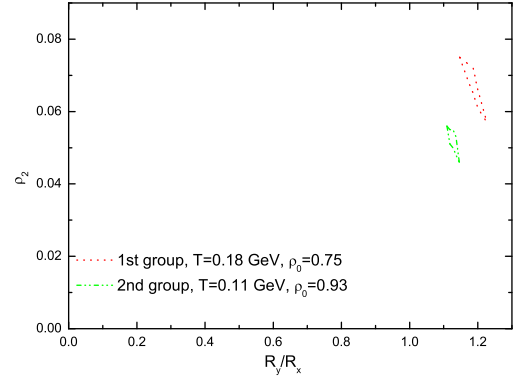} \caption{(Color online) Parameter contours for
$\rho_2$ and $R_{\rm y}/R_{\rm x}$ enclosing the regions of consistency
with RL blast-wave fits to $p_T$ spectra and $v_2$ in the thermal regime
for both freezeout groups. The temperature, $T$, and average radial-flow
rapidity, $\rho_0$, have been fixed at ($T,\rho_0$)=(180\,MeV, 0.75) and
(110\,MeV,0.93) for group-I and -II, respectively.}
\label{fig_contour}
\end{figure}
It is instructive to elaborate on the qualitative influence of the
fit parameters on the observables. As is well-known from more simple
blast-wave fits, the interplay of radial flow ($\rho_0$) and
temperature ($T$) can be disentangled by fits to hadrons with
different masses. With these 2 parameters fixed, an increase in
$\rho_2$ leads to a rapid increase in $v_2(p_T)$ at relatively large
$p_T$ but to rather small changes at low $p_T$.  On the contrary,
increasing $R_{\rm y}/R_{\rm x}$ results in a more uniform increase
of $v_2(p_T)$ over a wide range of $p_T$. As a result, a relatively
large $\rho_2$, combined with a relatively small eccentricity, i.e.,
$R_{\rm y}/R_{\rm x}$ close to unity, are favored in fitting the
data. The rather tight correlation between those two parameters can
be exhibited in pertinent contour plots indicating the ``allowed
regions" for obtaining an acceptable fit to data given their error
bars, see~Fig.~\ref{fig_contour}. The central values of the contours
for the two particle groups correspond to the values listed in
Tab.~\ref{tab:RLfit}. The contour for group-II hadrons is tighter
due to the larger number of data points for bulk particles. We first
note that the contours clearly exclude each other corroborating the
significant evolution that the collective flow field is undergoing
in the hadronic phase. More specifically, the anisotropy parameter
$R_{\rm y}/R_{\rm x}$ for group-II particles is smaller than for
group-I, which is in line with the general expectation that the
fireball becomes more spherical in the expansion. However, the flow
anisotropy $\rho_2$ extracted from the RL blast-wave fits favors a
decrease from hadronization (group-I) to kinetic decoupling of the
bulk (group-II). This is contrary to the expectation from
hydrodynamic expansion that $\rho_2=(\rho_{\rm x}^s-\rho_{\rm
y}^s)/2$ should increase if the in-plane acceleration is always
larger than the out-of-plane acceleration ($\rho_{\rm x}^s$ and
$\rho_{\rm y}^s$ are the surface-flow rapidities along the $x$- and
$y$-direction, respectively; recall
Eq.~(\ref{transverse_rapidity})). We cannot resolve this question
within the present blast-wave model. A full hydrodynamic simulation
could give more insight, possibly requiring substantial viscosity
effects in the later stages of the hadronic evolution.

To summarize this and the previous section, we have produced a
space-dependent parameterization of anisotropic flow fields which,
when implemented into a rather simple thermal blast-wave description
of locally equilibrated matter, provides a good overall fit to a
large body of hadron data in semi-central Au-Au data in the
low-$p_T$ regime. As a key ingredient we have introduced the well
established concept of sequential freeze-out including just two
groups but with clearly distinct flow fields.

\section{$\mathbf{\ket}$ Scaling of $\mathbf v_2$}
\label{sec_KET-scal}
A signature result of ideal hydrodynamical simulations of URHICs is
the mass splitting of the hadron $v_2$ when plotted as a function of
hadron $p_T$, with a  larger $v_2$ for the lighter particles at a
given $p_T$. On the other hand, if the hydrodynamic results for the
hadron $v_2$ are plotted versus $\ket$ (at a given freeze-out
temperature), the splitting nearly vanishes, but not entirely. In
fact, a small reverse mass ordering of $v_2$ is found instead, as
pointed out in Refs.~\cite{STAR_KET_scaling2008,Heinz2009}. In other
words, the hydro-induced mass splitting of $v_2(p_T)$ between light
and heavy particles is not large enough to render their $v_2(\ket)$
curves degenerate. The sequential freeze-out picture advocated above
may hold an explanation to this apparent discrepancy between
hydrodynamic freeze-out (at fixed temperature) and experimental data
(which exhibit ``perfect" $\ket$ scaling): Random thermal motion
tends to isotropize $p_T$ distributions and thus counteracts
(``dilutes") the collective motion in producing nonzero $v_2$. Thus
one should expect that a gap in freeze-out temperatures between,
say, (light) pions and (heavy) $\Omega$ baryons increases the mass
splitting in a $v_2(p_T)$ plot (the thermal motion of pions is
suppressed at lower temperatures), just enough to produce degeneracy
in a $v_2(\ket)$ representation. This appears to be a conspiracy of
circumstances that leads to the empirically observed $\ket$-scaling.
However, the underlying mechanism (largely different hadronic
rescattering cross sections) is general in the sense that it applies
to different centralities, system sizes and even collision energies,
presumably coupled with a rather general (local) freeze-out
criterion. It therefore could be a candidate for the origin of the
widely observed $\ket$ scaling. In the following, we quantitatively
test this idea by analyzing the scaling properties within our
sequential freezeout 2-source ansatz.

\begin{figure}[!t]
\hspace{4mm}
\includegraphics[width=\columnwidth
]{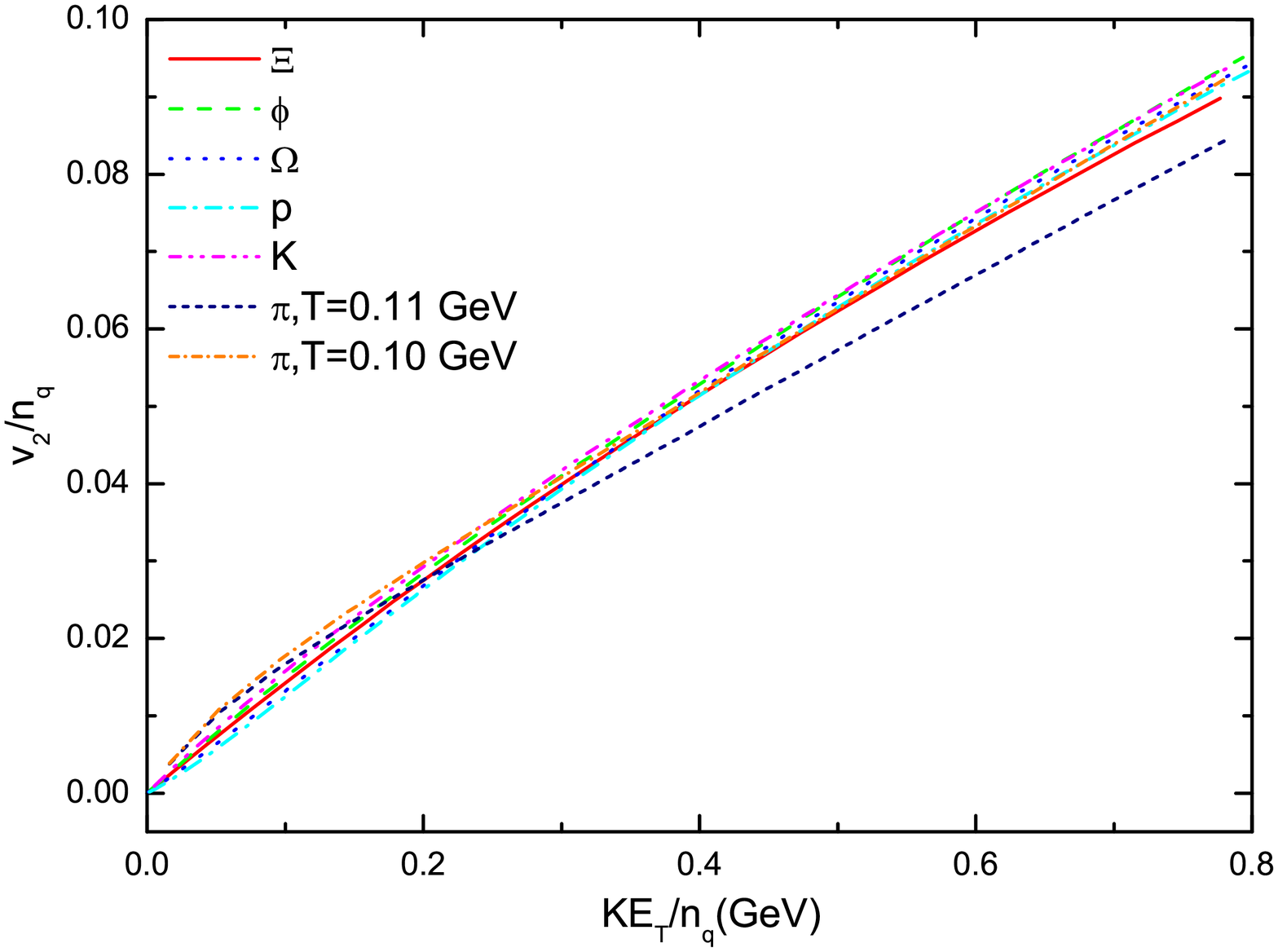}\caption{(Color online) $\ket$- and $n_q$-scaling
of $v_2$ for different hadrons obtained within the sequential
freeze-out scenario with blast-wave parameters given in Table
\ref{tab:RLfit}. The result of an slightly later freeze-out for
pions at temperature 100 MeV is also shown.} \label{fig_KET-scal}
\end{figure}

\begin{figure}[!t]
\hspace{-4mm}
\includegraphics[width=\columnwidth
]{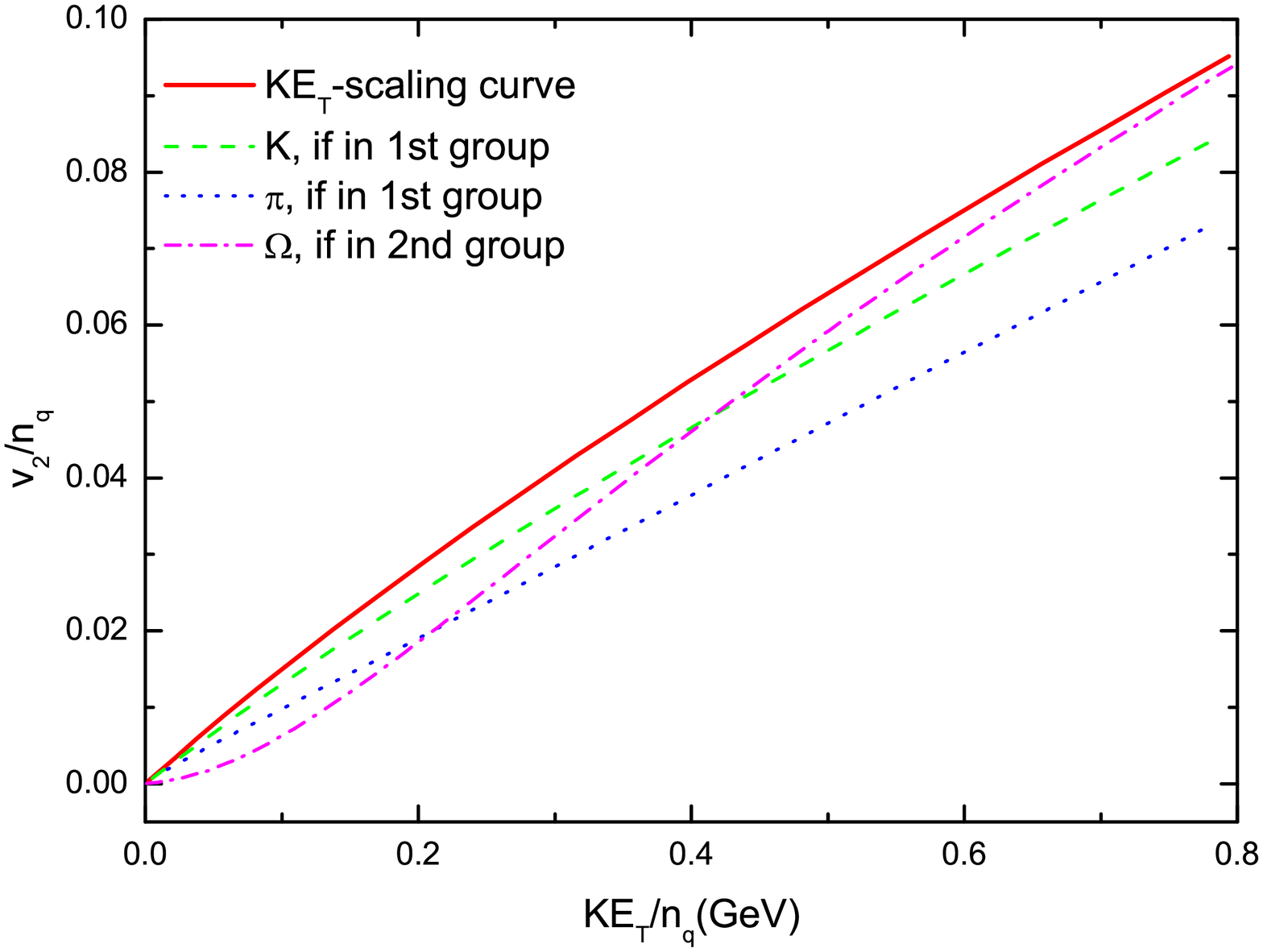}
\caption{(Color online) Comparison of the generic
$\ket$-scaling behavior of hadron-$v_2$ extracted from
Fig.~\ref{fig_KET-scal} (thick solid line) to results for
interchanging the freeze-out configurations for group-I and group-II
particles. Specifically, the $\Omega$ baryon (dash-dotted line) is
frozen out from the late source at $T_{\rm fo}=110$\,MeV, while
pions (dotted line) and kaons (dashed line) are frozen out from the
early source at $T_{c}=180$\,MeV (source parameters as quoted in
Tab.~\ref{tab:RLfit}).}.
\label{fig_KET-comp}
\end{figure}

With the RL flow-field parameterizations and freeze-out temperatures
for the two sources of group-I and group-II particles summarized in
Tab.~\ref{tab:RLfit} we proceed to display the $n_q$-scaled
hadron-$v_2$ as a function of the $n_q$-scaled hadron-$\ket$ in
Fig.~\ref{fig_KET-scal}. We find that all the hadrons-$v_2$ curves
follow a generic scaling very well (with slights deviations for the
pions, on which we comment below), very similar to experiment. This
agreement is, of course, not unexpected given that our results
follow from fits to data, together with the notion that the data
satisfy the scaling in the first place (or even led to its
discovery).  The nontrivial insight here is that our ``minimalistic"
approach of a sequential freeze-out scenario coupled with a rather
simple parameterization of an anisotropic collective flow field, is
sufficient to establish the scaling. We also recall that
$n_q$-scaling in the equilibrated low-$p_T$ region is not a direct
consequence of quark recombination, at least not for group-II
particles. To better exhibit the significance of the sequential
freeze-out we compare in Fig.~\ref{fig_KET-comp} the generic scaling
curve extracted from Fig.~\ref{fig_KET-scal} with $v_2$ results
where the (rather heavy) $\Omega$ baryon (group-I) is frozen out
from the source at $T_{\rm fo}=110$\,MeV (as sometimes done in
hydrodynamic calculations with a single freeze-out), and the (rather
light) pions and kaons are frozen out from the source at
$T_c=180$\,MeV (as advocated in recombination models that neglect
the hadronic phase). One finds a considerable breaking of both
$\ket$- and $n_q$-scaling. For the heavy particles, the increase in
collective flow at the low freeze-out temperature results in an
appreciable depletion of their $v_2$ at low $\ket$ (mass ordering
effect), thus spoiling their compatibility with scaling. On the
other hand, for pion and kaon freeze-out at higher temperature the
increased thermal motion in the local rest frame dilutes the
collective elliptic flow considerably and suppresses their $v_2$
below the scaling curve especially toward higher $\ket$.

Let us return to the deviations of the pion $v_2$ from the scaling
curve which become noticeable for $\ket/n_q>0.3$\,GeV in
Fig.~\ref{fig_KET-scal}. One may wonder whether feed-down
corrections from heavier particles ($\rho\to\pi\pi$,
$\omega\to3\pi$, $\Delta\to N\pi$, etc.) could improve the
situation. This question is addressed in detail in
Appendix~\ref{app_feeddown}, with the finding that the combined
effect of the decay pions increases the pion $v_2$ at $p_T>0.5$\,GeV
only a little (by a few percent). It turns out, however, that a
slight decrease of the pion freeze-out temperature from 110\,MeV to,
e.g., 100\,MeV increases the pion $v_2$ preferentially at higher
$\ket$, resulting in good agreement with the general scaling curve.
In doing so we have not performed a possibly small re-adjustment of
the flow parameters to attempt a more quantitative conclusion. The
key mechanism is again a further suppression of the random thermal
motion of the pions, thus augmenting the relative importance of the
collective motion. A 5-10\,MeV smaller freeze-out temperature is
certainly compatible with experimental data, and also not
unreasonable from the theoretical point of view. Pions have large
resonant cross sections on all major particle types of the bulk
(pion, kaon and nucleon), and their mass being the smallest makes it
plausible that their momentum spectra are frozen the latest (mass
ordering of thermal relaxation times).

\section{Summary and Conclusions}
\label{sec_concl}
In this paper we have analyzed the scaling properties of hadron
elliptic flow in ultrarelativistic heavy-ion collisions, which have
received considerable attention since the experimental data suggest
the existence of a universal behavior possibly related to a
collectively expanding partonic source. We have focused our
investigations on the low-$p_T$ regime ($p_T\lesssim 2$\,GeV) where
RHIC data (and hydrodynamic analysis thereof) support the assumption
of a locally equilibrated medium.

In the first part of this work we have pursued the earlier idea that
hadronization of the putative partonic system at RHIC proceeds
through coalescence of quarks and antiquarks. Its implementation in
the thermal regime requires a recombination model which obeys
4-momentum conservation and detailed balance to encode the correct
equilibrium limit. The Resonance Recombination Model (RRM) satisfies
these requirements. The flow field has been modeled by a blast-wave
ansatz with azimuthal asymmetries which give rise to space-momentum
correlations characteristic for a collectively expanding source.
Using RRM we have shown that an expanding parton source converts
into hadron spectra with the same collective source properties,
which, to our knowledge, has not been achieved in existing models of
quark coalescence. We have utilized this connection to extract quark
distributions just above the critical temperature for Au-Au
collisions at RHIC from measured $p_T$ spectra and elliptic flow of
multi-strange hadrons which are believed to undergo negligible
final-state interactions in the hadronic phase (group-I particles).

Bulk particles with large hadronic cross sections (e.g., pion, kaons
and nucleons) cannot be approximated by decoupling at the
hadronization transition. We have included these particles
(group-II) in our analysis by constructing another flow field at the
typical kinetic freeze-out temperature of 110~MeV, and fitting the
blast-wave parameters to their $p_T$ spectra and elliptic flow.
Group-II hadrons are therefore not really sensitive to quark
recombination. However, we find that the sequential freeze-out
picture with just two sources is well compatible with the
empirically observed kinetic-energy and constituent-quark number
scaling (where the latter is not directly linked to recombination).
The sequential freeze-out naturally corrects for a slight breaking
of $\ket$ scaling present in a hydrodynamic flow description for a
uniform freeze-out. In particular, a smaller temperature is
instrumental in suppressing the random thermal motion of the light
pions and in this way bring up their $v_2$ toward the universal
curve. In the thermal regime, CQNS of elliptic flow simply follows
from the universal {\em linear} curve resulting from $\ket$ scaling
(i.e., it is not directly linked to recombination).

An important task of future work will be an extension of our studies
to the intermediate-$p_T$ region. Here, the onset of the kinetic
regime occurs and thus off-equilibrium effects become important.
While this increases the sensitivity to quark degrees of freedom,
it also requires a more detailed knowledge of the reinteractions of
particles, both in the QGP and hadronic phase. Initial studies using
Langevin simulations for strange and heavy quarks have been
reported~\cite{Ravagli:2008rt}, but the scaling properties found
there clearly require a deeper understanding, including the effects
of baryon formation and a more microscopic description of the
elastic (as well as radiative) interactions in the QGP and hadronic
matter.

\vspace{0.3cm}

\acknowledgments

MH acknowledges helpful discussions with H.~van Hees, X.~Zhao and F.~Riek.
This work was supported by U.S. National Science Foundation (NSF) CAREER
Awards PHY-0449489 and PHY-0847538, by NSF grant PHY-0969394, by the
A.-v.-Humboldt Foundation, by the RIKEN/BNL Research Center, and by DOE
grant DE-AC02-98CH10886.

\appendix
\section{Feed-Down to Pions From Resonance Decays}
\label{app_feeddown}
\begin{figure}[!t]
\hspace{4mm}
\includegraphics[width=\columnwidth
]{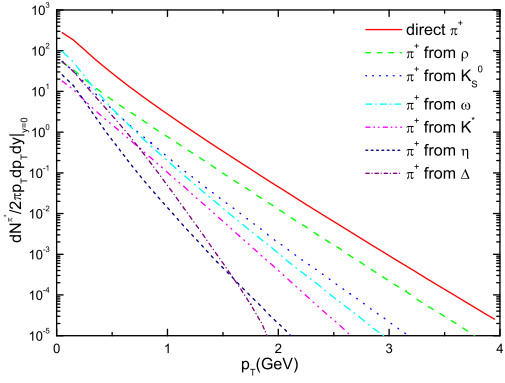} \caption{(Color online) Transverse momentum
spectra of pions from various resonance decays compared to direct
pions (solid line) at $T_{\rm fo}=110$\,MeV. The chemical potentials
for the parent particle are resonances are $\mu_{\rho}=180$\,MeV,
$\mu_{\Delta}=460$\,MeV, $\mu_{K^*}=250$\,MeV,
$\mu_{\omega}=270$\,MeV, $\mu_{\eta}=170$\,MeV, resulting from a
partial chemical equilibrium scenario in which chemical potentials
for stable particles are $\mu_{\pi}=90$\,MeV, $\mu_{K}=160$\,MeV and
$\mu_{N}=370$\,MeV.} \label{fig_pion-decay-pt}
\end{figure}

Hadron-resonance decays are known to increase pion yields
significantly for transverse momenta up to $p_T\simeq$
0.5-1~GeV~\cite{Heinz1991}. Therefore, it is mandatory to evaluate
the effect of the corresponding feed-down not only on the $p_T$
spectra but also on the elliptic flow of the
pions~\cite{Ko2004,Xu2004}. In particular, we want to understand how
feed-down affects the pion $v_2$ in the context of the
$\ket$-scaling for pion freeze-out at a temperature of $T_{\rm
fo}=110~{\rm MeV}$ where some deviation from protons and kaons in
group I has been identified (recall Fig.~\ref{fig_KET-scal}). The
total differential pion spectrum can be decomposed as
\begin{equation}
\frac{dN_\pi^{\rm tot}}{p_Tdp_Td\phi_pdy}=
\frac{dN_\pi^{\rm dir}}{p_Tdp_Td\phi_pdy}
+\sum_R\frac{dN_{R\rightarrow \pi}}{p_Tdp_Td\phi_pdy} \, ,
\label{pion_decay}
\end{equation}
where the summation is over all resonances $R$ with significant
branching fraction into final states containing pions. In our decay
simulation, we include $\rho$, $\omega$, $K^*$, $\Delta$, $\eta$ and
$K_0^S$ resonances, whose anisotropic momentum spectra, together
with that of direct pions, are obtained from the RL blast-wave
calculation with the same parameters as those for group II in
Tab.~\ref{tab:RLfit}. The resonance decays which contribute to the
final spectra occur at or after freeze-out. However, the absolute
number of the feed-down contribution, which is important to
determine the relative contribution to the total pion-$v_2$, depends
on the chemical potentials of the resonances at kinetic freeze-out.
In the interacting hadronic phase effective chemical potential need
to be introduced for hadrons stable under strong interactions (pion,
kaon, $\eta$, nucleon, etc.), as to conserve their total numbers
which are frozen at chemical
freeze-out~\cite{Bebie1992,Koch1997,Hung1998}. Part of this
procedure is to keep the strong resonances in relative chemical
equilibrium, e.g., $\mu_\rho=2\mu_\pi$ or $\mu_\Delta = \mu_N
+\mu_\pi$. At RHIC, the conservation of antibaryon number is of
particular importance in the overall chemistry of the hadronic
phase, and thus for the actual values of the effective chemical
potentials~\cite{Rapp:2000gy,Rapp:2002fc,Teaney:2002aj,Hirano2002,Kolb:2002ve}.
The values of the chemical potentials of the resonances at $T_{\rm
fo}$ are quoted in the caption of Fig.~\ref{fig_pion-decay-pt}; they
are taken from Ref.~\cite{Rapp:2002fc} where they are computed from
an isentropic evolution starting at chemical freeze-out at
$T=180$\,MeV (consistent with our parameters for group-1
freeze-out). Note that the $K_0^S$ decays through weak interactions
but carries its own chemical potential (as does the $\eta$).

The $p_T$-spectra of pions from various resonance decays are
compiled in Fig.~\ref{fig_pion-decay-pt}. The decay pions generally
have steeper $p_T$-spectra than directly produced pions, except for
pions from $\rho$ decays, similar to the observations made in
Refs.~\cite{Ko2004,Xu2004}. This is essentially a phase-space
effect: pions from resonances with a small $Q$-value (the difference
of the resonance mass and the sum over masses of its constituent
quarks) tend to have a steeper $p_T$ spectrum. The phase-space also
affects the $v_2$ of decay pions. The momenta of pions from
resonances with large $Q$ values (like the $\rho$) tend to dilute
the $v_2$ more due to the random orientation of the decay momentum,
most notably at low momenta (where the decay momentum in the
parent's rest frame is comparable to the parent momentum). On the
other hand, pions from resonances with small $Q$ values (like the
$\Delta$) are essentially aligned with the parent and thus preserve
the elliptic flow of the parent. For those pions, the $v_2$ could be
larger than that of the parent at the same $p_T$, since the momentum
of the resonance is shared by several daughters. However, since the
dominant source of secondary pions are $\rho$ decays, the total
pion-$v_2$ might not be much affected by resonance decays.

\begin{figure}[!t]
\hspace{4mm}
\includegraphics[width=\columnwidth
]{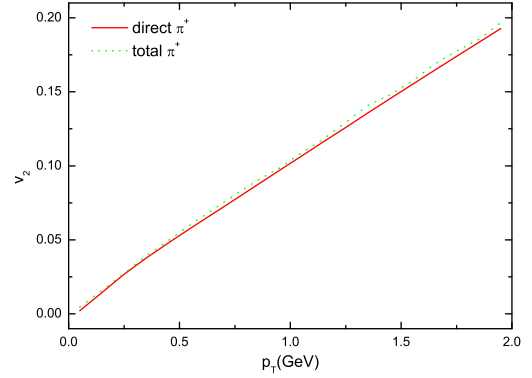} \caption{(Color online) Total pion-$v_2$ (dashed
line) compared to direct pion $v_2$ (solid line) at $T_{\rm
fo}=0.11~{\rm GeV}$.} \label{fig_pion-decay-v2}
\end{figure}
The comparison of the total pion-$v_2$ to the direct-pion-$v_2$
confirms this expectation, see Fig.~\ref{fig_pion-decay-v2}.
Therefore, feed-down contributions are not sufficient to resolve the
discrepancy between the direct-pion $v_2$ and the generic scaling
curve in the $\ket$-dependence of $v_2$. In comparison to previous
estimates we note that in Ref.~\cite{Ko2004} a larger enhancement of
the total pion-$v_2$ over the direct-pion $v_2$ was reported. In
that work, the resonance decays were evaluated at $T_c$; subsequent
hadronic rescattering was neglected and the direct pions were
computed from a quark coalescence model which does not conserve
energy. Both of these approximations may be unreliable for pions at
small $p_T$, due to their large interaction cross section and
binding energy, respectively. The influence of resonance decays on
pion elliptic flow was also studied in
Ref.~\cite{Steinberg_KET_scaling2008}, where the authors assumed
that direct pions follow the $KE_T$-scaling and then found the
scaling still holds at the $10\%$ level when resonances are
included.


\begin{thebibliography}{99}

\bibitem{RHIC2005}
  J. Adams {\it et al.} (STAR Collaboration), Nucl. Phys. {\bf
  A757}, 102 (2005); K. Adcox {\it et al.} (PHENIX Collaboration), Nucl.
  Phys. {\bf A757}, 184 (2005); I. Arsene {\it et al.} (BRAHMS
  Collaboration), Nucl. Phys. {\bf A757}, 1 (2005); B. B. Back {\it et
  al.} (PHOBOS Collaboration), Nucl. Phys. {\bf A757}, 28 (2005).
\bibitem{v2review2009}
  P. Sorensen, arXiv:0905.0174v3 [nucl-ex]; S. A. Voloshin, A. M.
  Poskanzer, and P. Snellings, arXiv:0809.2949v2 [nucl-ex].
\bibitem{RHICv2}
  J. Adams {\it et al.} (STAR Collaboration and STAR-RHIC Collaboration),
  Phys. Rev. C {\bf 72}, 014904 (2005).
\bibitem{shuryak}
  D. Teaney, J. Lauret, and E.V. Shuryak, Phys. Rev. Lett. {\bf 86}, 4783 (2001);
  D. Teaney, J. Lauret, and E.V. Shuryak, arXiv:nucl-th/0110037.
\bibitem{Heinz20032004}
  U. Heinz, arXiv:hep-ph/0407360v1; P. F. Kolb and U. Heinz,
  arXiv:nucl-th/0305084v2.
\bibitem{Heinz2000}
  P. F. Kolb, J. Solfrank, and U. Heinz, Phys. Rev. C {\bf 62},
  054909 (2000).
\bibitem{Heinz2001}
  P. Huovinen {\it et al.}, Phys. Lett. {\bf B503}, 58 (2001).
\bibitem{Hirano2002}
  T. Hirano and K. Tsuda, Phys. Rev. C {\bf 66}, 054905 (2002); T.
  Hirano and M. Gyulassy, Nucl. Phys. {\bf A769}, 71 (2006).
\bibitem{1st_scaling}
  S. A. Voloshin, Nucl. Phys. {\bf A715}, 397c, (2003).
\bibitem{coalescencemodels}
  D.\ Molnar and S.\ A.\ Voloshin, Phys.\ Rev.\ Lett.\ {\bf 90}, 202303
  (2003); R.\ Fries {\it et al.}, Phys.\ Rev.\ C {\bf 68}, 044902
  (2003); V.\ Greco, C.\ M. Ko, and P.\ Levai, Phys.\ Rev.\ C {\bf 68},
  034904 (2003); R.\ Fries, V.\ Greco, and P.\ Sorensen, Ann.\ Rev.\
  Nucl.\ Sci.\ {\bf 58}, 177 (2008).
\bibitem{Pratt:2004zq}
  S.~Pratt and S.~Pal, Nucl.\ Phys. {\bf A749}, 268 (2005); D.
  Molnar, arXiv: nucl-th/0408044v2; V. Greco and C. M. Ko, arXiv:
  nucl-th/0505061v2.

\bibitem{Phenix_KET_scaling2007}
  A. Adare {\it et al.} (PHENIX Collaboration), Phys. Rev. Lett. {\bf
  98}, 162301 (2007);  B. I Abelev {\it et al.} (STAR Collaboration), Phys. Rev. C {\bf
  75}, 054906 (2007).
\bibitem{STAR_KET_scaling2008}
  B. I. Abelev {\it et al.} (STAR Collaboration), Phys. Rev. C {\bf
  77}, 054901 (2008).
\bibitem{Steinberg_KET_scaling2008}
  L. A. Linden Levy {\it et al.}, Phys. Rev. C {\bf 78}, 044905
  (2008).
\bibitem{Lacey2006}
  R. A. Lacy and A. Taranenko, arXiv:nucl-ex/0610029v3.
\bibitem{Ma2009}
  J. Tian {\it et al.}, Phys. Rev. C {\bf 79}, 067901 (2009).
\bibitem{Jia2007}
  J. Jia and C. Zhang, Phys. Rev. C {\bf 75}, 031901(R) (2007).

\bibitem{Ravagli:2007xx}
  L.~Ravagli and R.~Rapp, Phys.\ Lett. {\bf B655}, 126 (2007).
\bibitem{Cassing}
  W. Cassing and E. L. Bratkovskaya, Phys. Rev. C {\bf 78}, 034919 (2008).

\bibitem{Ravagli:2008rt}
  L.~Ravagli, H.~van Hees, and R.~Rapp, Phys.\ Rev.\  C {\bf 79}, 064902 (2009);

\bibitem{vanHees:2005wb}
  H.~van Hees, V.~Greco, and R.~Rapp,
  Phys.\ Rev.\  C {\bf 73}, 034913 (2006).
\bibitem{Olga_sequential}
  O. Barannikova (for STAR collaboration), arXiv:nucl-ex/0403014.
\bibitem{Xu2009}
  B. Mohanty and N. Xu, J. Phys. G: Nucl. Part. Phys. {\bf 36},
  064002, (2009).
\bibitem{more_coalescence}
  R. J. Fries {\it et al.}, Phys. Rev. Lett. {\bf 90}, 202303
  (2003); V. Greco, C. M. Ko, and P. Levai, Phys. Rev. Lett. {\bf
  90}, 202302 (2003); R. C. Hwa and C. B. Yang, Phys. Rev. C {\bf
  67}, 034902 (2003).
\bibitem{Lisa2004}
  F. Retiere and M. A. Lisa, Phys. Rev. C {\bf 70}, 044907 (2004).
\bibitem{Bjorken1984}
  J. D. Bjorken, Phys. Rev. D {\bf 27}, 140 (1983);
  R. C. Hwa, Phys. Rev. D {\bf 10}, 2610 (1974).
\bibitem{Heinz1993}
  E. Schnedermann, J. Solfrank, and U. Heinz, Phys. Rev. C {\bf 48},
  2462 (1993).
\bibitem{Hwa_Yang}
  R. C. Hwa and C. B. Yang, Phys. Rev. C {\bf 75}, 054904 (2007).
\bibitem{STAR_phi}
  B. I. Abelev {\it et al.} (STAR Collaboration), Phys. Rev. Lett.
  {\bf 99}, 112301 (2007).
\bibitem{STAR_multistrange_pT}
  J. Adams {\it et al.} (STAR Collaboration), Phys. Rev. Lett. {\bf 98}, 062301
  (2007).
\bibitem{STAR_phi_new}
B. I. Abelev {\it et al.} (STAR
  Collaboration), Phys. Rev. C {\bf 79}, 064903 (2009).
\bibitem{HZHuang}
  H. Z. Huang, J. Phys. G: Nucl. Part. Phys. {\bf 36}, 064008
  (2009); J. H. Chen {\it et al.}, Phys. Rev. C {\bf 78}, 034907
  (2008).
\bibitem{Rapp:2002fc}
  R.~Rapp,
  Phys.\ Rev.\  C {\bf 66}, 017901 (2002).
\bibitem{Rapp:2009yu}
R. Rapp, J. Wambach, and H. van~Hees, Landolt B\"ornstein
(Springer), New Series {\bf I/23-A}, 4-1 (2010); arXiv:0901.3289
[hep-ph].
\bibitem{STAR_proton}
  B. I. Abelev {\it et al.} (STAR Collaboration), Phys. Rev. Lett.
  {\bf 97}, 152301 (2006).
\bibitem{kaon_pT}
  S. S. Adler {\it et al.} (PHENIX Collaboration), Phys. Rev. C {\bf 69}, 034909
  (2004).
\bibitem{Heinz2009}
  U. Heinz, arXiv:0901.4355 [nucl-th].
\bibitem{Heinz1991}
  J. Sollfrank, P. Koch, and U. Heinz, Z. Phys. C {\bf 52}, 593
  (1991).
\bibitem{Ko2004}
  V. Greco and C. M. Ko, Phys. Rev. C {\bf 70}, 024901 (2004).
\bibitem{Xu2004}
  X. Dong {\it et al.}, Phys. Lett. {\bf B597}, 328 (2004).
\bibitem{Bebie1992}
  H. Bebie {\it et al.}, Nucl. Phys. {\bf B378}, 95 (1992).

\bibitem{Koch1997}
  C. Song and V. Koch, Phys. Rev. C {\bf 55}, 3026 (1997).

\bibitem{Hung1998}
 C.M.~Hung and E.V.~Shuryak, Phys. Rev. C {\bf 57}, 1891 (1998).
\bibitem{Kolb:2002ve}
  P.~F.~Kolb and R.~Rapp,
  Phys.\ Rev.\  C {\bf 67}, 044903 (2003).

\bibitem{Rapp:2000gy}
  R.~Rapp and E.V.~Shuryak,
  Phys.\ Rev.\ Lett.\  {\bf 86}, 2980 (2001)



\bibitem{Teaney:2002aj}
  D.~Teaney,
  arXiv:nucl-th/0204023.




\end{thebibliography}
\end{document}